\documentclass[a4paper,twocolumn,showpacs,amsmath,pra,amssymb]{revtex4-1} 
\usepackage[colorlinks=true,urlcolor=blue,citecolor=blue,linkcolor=blue]{hyperref}
\usepackage[T1]{fontenc}
\usepackage[latin9]{inputenc}
\usepackage{times}
\usepackage{color}
\usepackage{xspace}
\usepackage{amssymb,amsmath}
\usepackage{amsbsy}
\usepackage{graphicx}
\usepackage{float}
\usepackage[T1]{fontenc}
\usepackage{pslatex}
\usepackage{contour}
\usepackage{color}
\usepackage[usenames,dvipsnames]{xcolor}
\newcommand{\ket}[1]{|#1\rangle}

\definecolor{raw}{RGB}{255,177,100}
\definecolor{filtered}{RGB}{153,0,51}
\begin{document}
\title{Reducing number fluctuations of ultracold atomic gases via dispersive interrogation}
\author{Bianca J. Sawyer}
\author{Amita B. Deb}
\author{Thomas McKellar}
\author{Niels Kj{\ae}rgaard}\email{nk@otago.ac.nz}
\affiliation{Jack Dodd Centre for Quantum Technology, Department of Physics, University of Otago, Dunedin, New Zealand.}
\date{\today}
\begin{abstract}
We have used dispersive laser probing to follow the central density evolution of a trapped atomic cloud during forced evaporative cooling.  This was achieved in a heterodyne detection scheme. We propose to use this as a nondestructive precursor measurement for predicting the atom number subsequent to evaporation and provide a simple experimental demonstration of the principle leading to a conditional reduction of classical number fluctuations.
\end{abstract}
\pacs{37.10.Gh, 67.85.-d, 42.50.Nn}
\maketitle
Ultracold gases, notably Bose-Einstein condensates (BECs), have received considerable interest in recent years as a source for metrological applications such as atom interferometers \cite{RevModPhys.81.1051}. In this context, optical dispersive measurements have been put forward as a method for reducing condensate number fluctuations in feedback protocols \cite{Wiseman2001Reducing,Szigeti2010} and as a method for quantum state engineering of atomic ensembles \cite{Kuzmich2007Atomic}. The very act of measuring the phase shift of light passing through a cloud of atoms can define a collective pseudo-spin component beyond the standard quantum limit while retaining a significant atomic coherence. Indeed, several recent experiments have reported metrologically relevant measurement induced spin squeezing (see, e.g., \cite{Ma2011} for a review).

Outside the realm of measurements and stabilization at the quantum level, dispersive detection techniques have also found interesting applications. Phase contrast imaging has for instance been applied to film the formation of magnetic domains in spinor BECs \cite{Sadler2006}.  Furthermore, dispersive probing was used to follow breathing \cite{Petrov2007} and center-of-mass oscillations \cite{Kohnen2011} for trapped atomic clouds, real time recording of Rabi oscillations \cite{Chaudhury2006,Windpassinger2008,Bernon2011}, and the detection and correction of random rotations of a collective pseudo-spin \cite{2012arXiv1207.3203}.
In the context of the present paper, it is of notable interest that dispersive probing holds the potential to increase the duty cycle of a repeated experiment, which has direct metrological impact. For example, ``nondestructive'' measurements have been employed to reduce the Dick noise of an optical lattice clock by minimizing the dead time of the clock cycle \cite{Lodewyck2009Nondestructive}.
\begin{figure}[b!]
\begin{center}
    \includegraphics[width=\columnwidth]{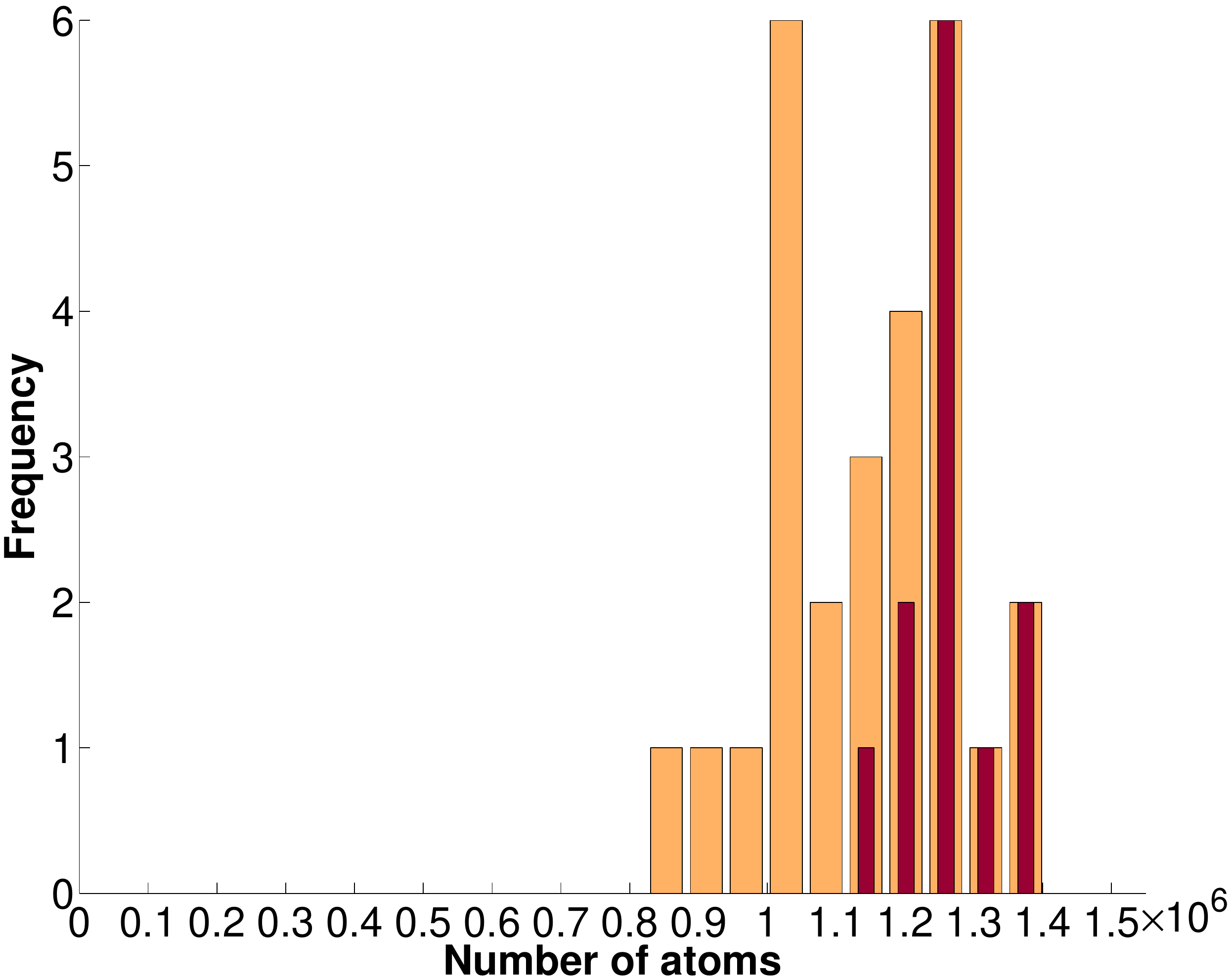}
\caption{(Color online) Histogram for number of atoms as observed in 23 consecutive experimental runs (light orange). Superimposed is the resulting distribution after filtering according to a dispersive precursor measurement (dark red). \label{fig:Hist2}}
\end{center}
\end{figure}
\begin{figure*}[t!]
\begin{center}
    \includegraphics[width=\textwidth]{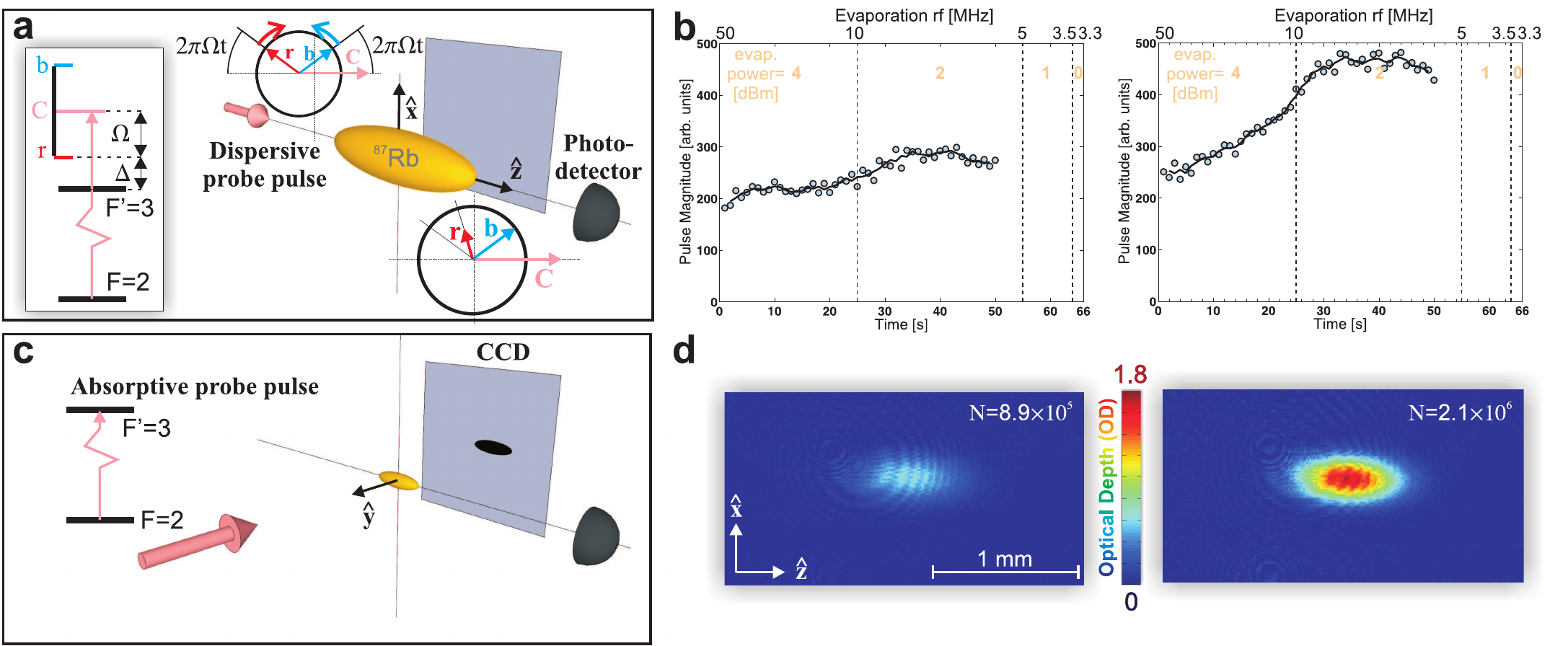}
\caption{(Color online) Concept of measurement scheme. (a) Dispersive probing: during the evaporative cooling stage the atoms are probed using an off-resonant laser light frequency triplet (inset). The phase-shift and Faraday rotation of the red sideband (r) upon passage through the sample is represented by phasor diagrams for polarization states parallel to the carrier (C) in a frame co-rotating with \textbf{C}. The photodetector experiences a beat note at frequency $\Omega$ as a result of the rotation and shortening of \textbf{r}. (b) The time evolution in the dispersive signal for a low (left) and a high (right) number sample. (c) At the end of the evaporation atoms are probed using conventional destructive absorption imaging. (d) Absorption images acquired at 4.5~ms time-of-flight, corresponding to the low (left) and high (right) number ($N$) samples followed in (b), respectively.\label{fig:concept}}
\end{center}
\end{figure*}

In this Brief Report, we demonstrate the reduction of (classical) number fluctuations of a magnetically trapped atomic gas after forced rf evaporative cooling to the ultracold domain. By using off-resonant probe laser light to measure the peak atomic column density near the beginning of the evaporation sequence we can predict with high likelihood the number of atoms  at the end of the sequence. Conditional on the outcome of the dispersive precursor measurement, it becomes possible to interrupt the experimental cycle at an early stage in the case of an unfavorable starting point, thereby saving precious experimental uptime.

Our experimental apparatus and its operation is standard \cite{Lewandowski2003}. Initially, a $\rm ^{87}Rb$ magneto-optical trap (MOT) is loaded, followed by a compressed MOT and molasses stage. The atoms are optically pumped to the $\ket{F=2, m_F=2}$ hyperfine substate and transferred into a magnetic quadrupole trap mounted on a motorized translation stage. This stage shuttles atoms to an ultra-high vacuum science cell where they are loaded into a ``clip-style'' Ioffe-Pritchard trap \cite{Ernst1998b}, characterized by axial and radial trapping frequencies of $\omega_a=2\pi\times 16$~Hz and $\omega_r=2\pi\times 153$~Hz, respectively. Finally, forced rf evaporation \cite{Ketterle1996} cools our sample to a temperature of $\rm\sim3~\mu K$.

In many applications using ultracold atomic clouds, a stable final atom number from run to run is desirable. Figure~\ref{fig:Hist2} shows a typical distribution in final atom numbers achieved in our experimental setup when continuously running at a 140~s duty cycle which allows for evaporation, the magnetic field coils to cool down, and the MOT to build up to a saturated level. The overall number distribution has a mean value of $1.15\times10^6$ atoms and a standard deviation of $1.4\times 10^5$. Obviously, the spread in numbers could be reduced by using a protocol that would discriminate against initial samples likely to result in a significant deviation from some target value. Figure \ref{fig:concept} illustrates how this can be achieved via dispersive interrogation of the atoms using the technique of fm spectroscopy \cite{Bjorklund1983,Lye1999}.

Our dispersive probing scheme is similar to that of Ref.~\cite{Bernon2011}. The  probe beam is generated with a diode laser and blue-detuned from the $\rm ^{87}Rb$ D2 line $F=2\rightarrow F'=3$ transition by $707\,\rm{MHz}$. An electro-optic phase modulator imprints a red (r) and a blue (b) sideband shifted by $\Omega=500\,\rm{MHz}$ from the carrier (C). This frequency triplet then passes through an acousto-optic modulator, used for pulsing the probe beam, which shifts the frequencies down by 80~MHz. The probing sideband (i.e. the red sideband) is hence blue-detuned by $\Delta=127$~MHz from the $F=2\rightarrow F'=3$ transition [see inset panel of Fig.~\hyperref[fig:concept]{\ref*{fig:concept}(a)}]; it contains $7\%$ of the total beam power. The dispersive probe beam propagates along the axial direction ($z$-axis) of our atomic sample and is centered on it radially, as illustrated in Fig.~\hyperref[fig:concept]{\ref*{fig:concept}(a)}. The probe beam has a waist of $w_0 = 30\,\rm{\mu m}$ and is linearly polarized along the $x$-axis, perpendicular to the 2.1~G bias field (pointing in the $z$-direction) at the magnetic trap center. On passage through the atomic sample the probing sideband incurs a dispersive phase shift and its polarization will rotate as a result of the Faraday effect \cite{Labeyrie2001}, both dependent upon the number of atoms in the $\ket{F=2, m_F=2}$ state inside the probe volume (the light-atom interaction for the carrier and upper sideband, being detuned by $627~\rm{MHz}$ and $1127~\rm{MHz}$, respectively, from the $2\rightarrow 3'$ resonance, is negligible in comparison). Consequently, an optical beat signal is generated at the modulation frequency of $\Omega=500~\rm{MHz}$, which we detect with a high speed biased photodetector. The resulting signal is amplified, demodulated at $500\,\rm{MHz}$, lowpass filtered at 100~kHz, and sampled using a digitizer in $200~\rm{\mu s}$ time segments around the probe pulses, having a duration of $20~\rm{\mu s}$.

To characterize our scheme, we probe our atomic sample, while evaporating, at a repetition rate of $1\,\rm{Hz}$, using 50 light pulses. The total optical power of the probe beam is $10\,\rm{\mu W}$ and the probing sideband thus contains $\sim5.5\times10^7$ off-resonant photons per pulse. The evaporation sequence consists of a series of four rf sweeps (linear in time) using (relative) drive powers as indicated in Fig.~\hyperref[fig:concept]{\ref*{fig:concept}(b)}.  At the end of evaporation the ultracold cloud is probed using standard absorption imaging \cite{Ketterle1999} with a resonant probe beam propagating along the $y$-axis as illustrated in \hyperref[fig:concept]{\ref*{fig:concept}(c)}. From this we determine the total number of atoms. We terminate the dispersive probe pulse train after the first 50 seconds of evaporation as the fragility of the atomic cloud increases substantially in the final two evaporation ramps and destructivity becomes significant; when probing for the full 66~s of the evaporation sequence, the sample suffers a $\sim60$\% loss in final atom number. For our laser parameters and at the center of the beam (i.e, at the peak intensity) the Kramers-Heisenberg formula \cite{Loudon2003,*Cline1994} stipulates a $\sim35$\% loss of atoms from the magnetically trapped $\ket{F=2, m_F=2}$ state during a probe pulse due to spontaneous, inelastic Raman scattering of photons from the probing side band as well as the carrier light \cite{Lye2004}. The remaining, trapped atoms (again, taken at the peak intensity of the probe light) would have scattered (elastic Rayleigh process) 5~photons on average - each heating an atom by twice a photon recoil energy ($E_{\rm rec/k_B}\sim180~nK$) \cite{Miller1993}. Hence a probe pulse of the present parameters will significantly affect an ultracold sample with a radial extent $r\lesssim w_0$. At the time of our final probe pulse the atomic sample has a temperature of ${\rm \gtrsim 30~\mu K}$ and a radius of $r{\rm \gtrsim 80~\mu m}$ , i.e., still reasonably robust against probing \cite{Note1}. The associated atomic loss and heating from probing is completely negligible at the beginning of the evaporation, where we are dealing with millimeter-sized samples ($r\gg w_o$) containing billions of atoms at temperatures much higher than the recoil temperature for spontaneous emission.

A processed data set $\{d_{i}\}$ for the series of pulses $i=1...50$ is obtained by integrating over the width of each pulse and subtracting from this a background value. Figure~\hyperref[fig:concept]{\ref*{fig:concept}(b)} shows examples of the sequential development in $d_i$ for two experimental runs corresponding to a high and a low number atomic sample. Figure~\hyperref[fig:concept]{\ref*{fig:concept}(d)} shows the corresponding absorption images as acquired at the end of evaporation. $\{d_i\}$ is a measure of the central column density of the sample along the $z$-axis as long as $w_0$ is smaller than the sample radius. The observed behavior in Fig.~ \hyperref[fig:concept]{\ref*{fig:concept}(b)}  is explained well by a simple theoretical model based on the spatial sample-probe overlap integral \cite{Windpassinger2008a}. Initially, $d_i$ is observed to increase as a result of evaporation and an increase in optical depth (OD), but eventually levels off and decreases as the radial extent of the atomic sample becomes comparable to and smaller than $w_0$.

\begin{figure}[t!]
\begin{center}
    \includegraphics[width=\columnwidth]{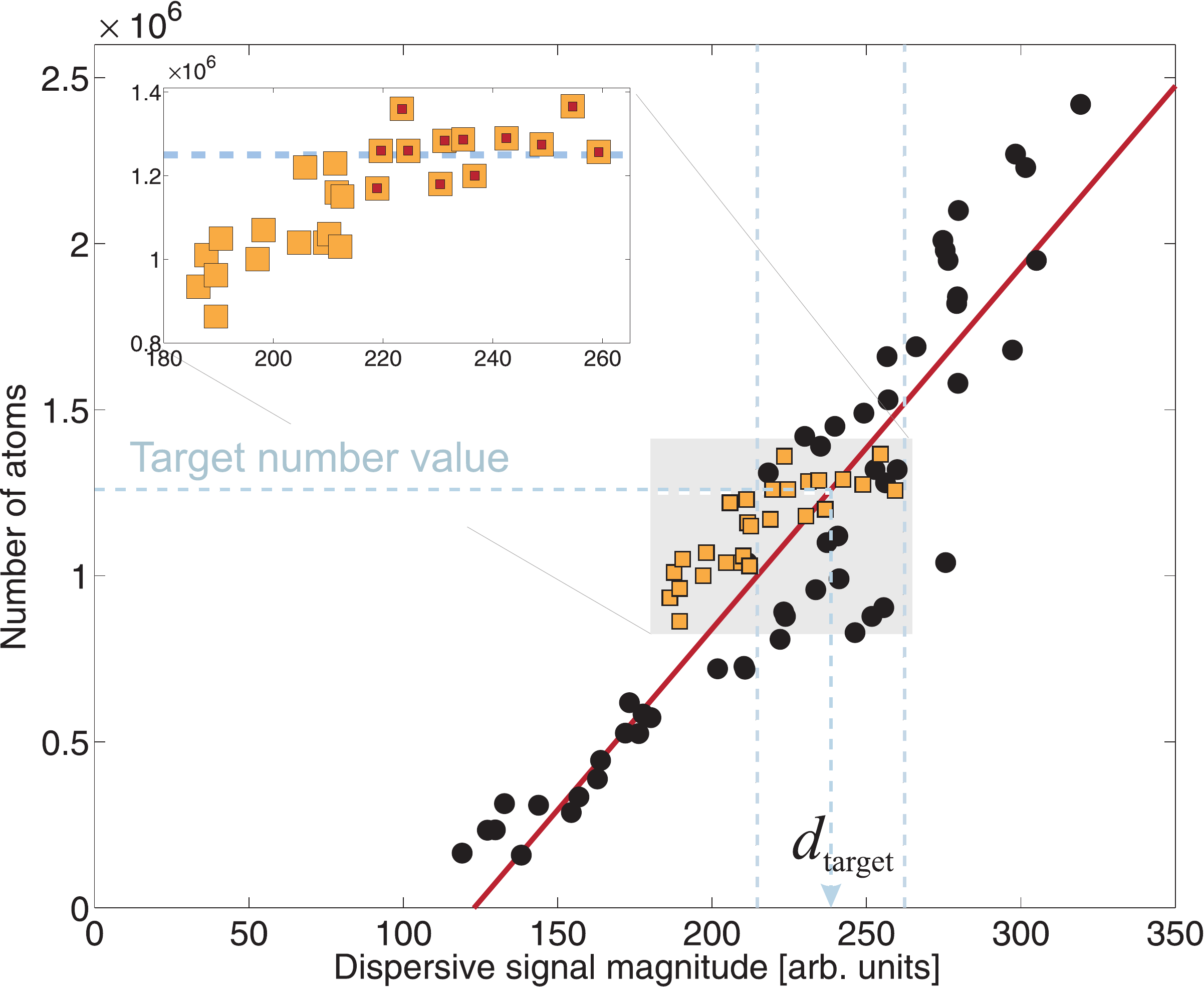}
\caption{(Color online) Number of atoms $a(k)$ as measured in absorption imaging at the end of the evaporation sequence versus the dispersive interrogation pulse $d_{10}(k)$ ({\large$\bullet$}); red line is a least squares fit to the data, which is used to define $d_{\rm target}$ from a target atom number. The dataset corresponding to Fig.~\ref{fig:Hist2} ({\scriptsize\textcolor{raw}{$\blacksquare$}}) is decimated ({\scriptsize\textcolor{filtered}{$\blacksquare$}}) by excluding points outside a 10\% band around $d_{\rm target}$.\label{fig:corrplot}}
\end{center}
\end{figure}

For our characterization, we acquired data in 53 consecutive experimental runs (designated by $k$) where the atom numbers were deliberately varied to yield a significant dynamical signal range. This was achieved by adjusting the optical power of the MOT laser beams, resulting in a variation in the magnetic trap loading. In Fig.~\ref{fig:corrplot} we plot the final atom number $a(k)$ as inferred from absorption imaging versus the dispersive signal strength for the $10^{\rm{th}}$ pulse $d_{10}(k)$. Previous investigations have shown a close to linear relation beween MOT levels and final atom number for a fixed evaporation sequence \cite{Papp2007}. Similarly, we find that the atom number at the end of the evaporation sequence scales linearly with the dispersive signal 10~s into the evaporation sequence. As a measure of the correlation we compute the data sample Pearson correlation coefficient

\begin{figure}[b!]
\begin{center}
\includegraphics[width=\columnwidth]{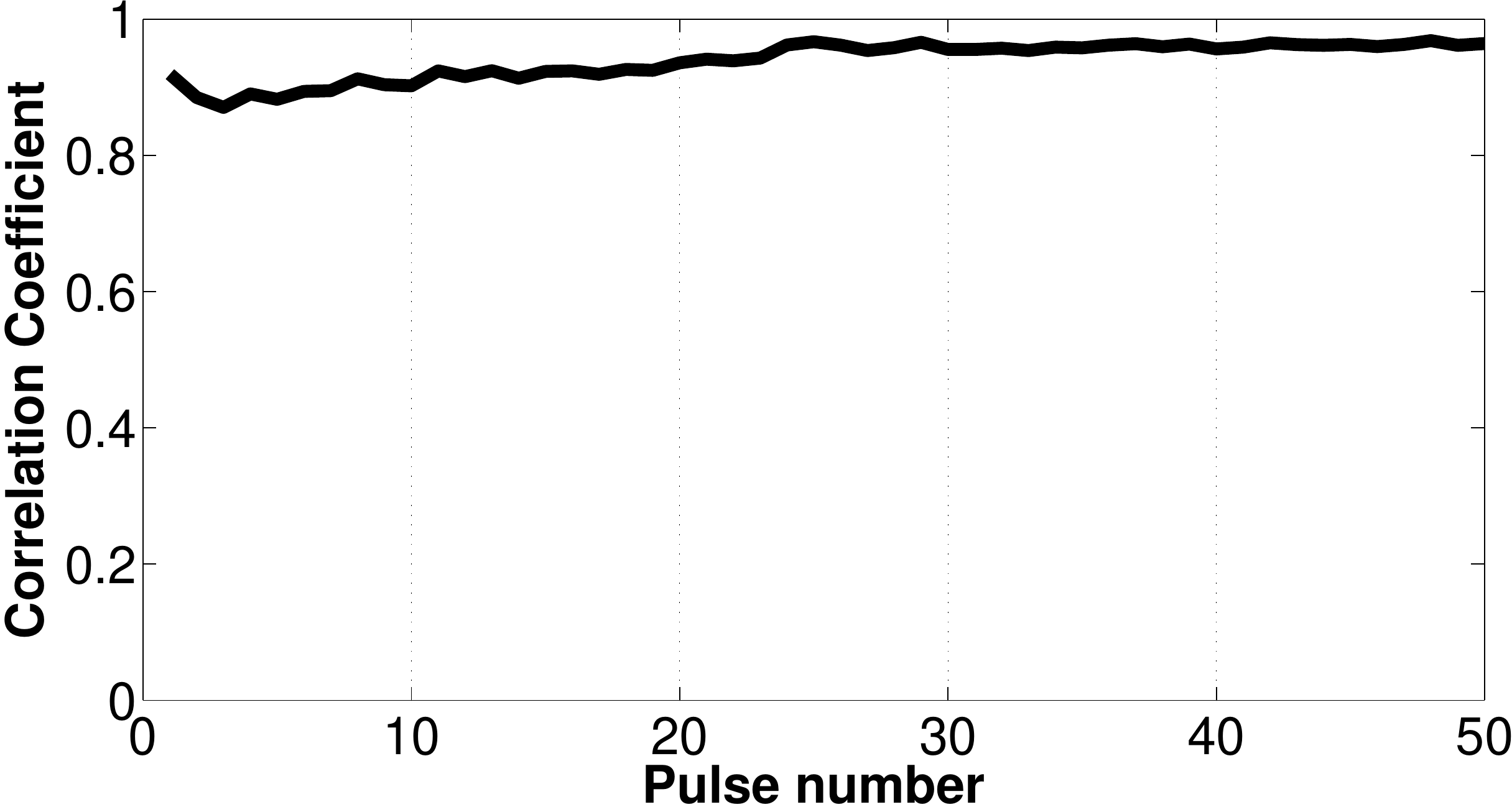}
\caption{Correlation coefficient $r_i$ versus pulse number $i$. \label{fig:corrcoef}}
\end{center}
\end{figure}
\begin{equation}\label{eqcorr coeff}
r_i=\frac{\sum_{k=1}^N [a(k)-\langle{a}\rangle][d_i(k)-\langle{d_i}\rangle]}{\sqrt{\sum_{k=1}^N [a(k)-\langle{a}\rangle]^2}\sqrt{\sum_{k=1}^N [d_i(k)-\langle{d_i}\rangle]^2}}.
\end{equation}
 Figure~\ref{fig:corrcoef} shows the progression of $r_i$ with dispersive probe pulse number $i$. As would be expected, the correlation coefficient steadily increases with $i$ as the time separation between $d_i(k)$ and $a(k)$ decreases. However, even for low values of $i$ we note a significant correlation. For example, for the data of Fig.~\ref{fig:corrplot} (based on $i=10$) we obtain $r_{10}=0.90$

 The high degree of correlation immediately implies that a dispersive precursor measurement can be exploited to reduce the number fluctuations in the final sample. Whilst collecting samples for the number distribution in Fig.~\ref{fig:Hist2} we simultaneously acquired the complementary dispersive data. We now choose the magnitude of the dispersive signal at the 10 pulse mark as a precursor measurement for predicting the final atom number and we take $1.25\times 10^6$ atoms as the target value for our filtering process. The target number value in turn defines a corresponding dispersive signal $d_{\rm target}$ [defined by a linear fit to $\{d_{10}(k),a(k)\}$ --- See Fig.~\ref{fig:corrplot}]. By excluding experimental realizations with a precursor value outside a 10\% band around $d_{\rm target}$ we get the filtered data shown in dark red in Fig~\ref{fig:Hist2} and the inset of Fig.~\ref{fig:corrplot}, superimposed on the raw data. The filtered number distribution has a mean of $1.27\times 10^6$ and has a width characterized by a standard deviation of $6.1\times 10^4$ which is a reduction of factor of $\gtrsim2$ in comparison to the original unfiltered distribution.

In conclusion, we have demonstrated that the information gained by probing a trapped atomic sample at the beginning of the evaporation sequence using a reasonably simple nondestructive, dispersive heterodyne scheme displays a significant correlation with the final resulting atom number. This may be used in a preselecting protocol to achieve consistent, high number samples. While the precursor interrogation pulse only samples a fraction of the cloud via its column density at a particular point it provides a quite robust quality measure. Such complimentary diagnostics tools may prove valuable as cold atoms experiments move towards applications outside the laboratory \cite{Geiger2011}.
We also note that nondestructive probing of an atomic sample during evaporative cooling may open up paths for more efficient adaptive optimization procedures than conventional and tedious algorithms using absorption imaging \cite{Lewandowski2003}. In particular, stochastic optimization of BEC production using genetic algorithms \cite{Rohringer2011,*Rohringer2008}, presently based on absorption images as the sole input, may benefit from the additional information gained. Along these lines, we point out that field programmable gate arrays would provide an ideal solution for acquiring and processing the dispersive signal and actuate on the sample in a feedback loop.

Minimally destructive detection techniques were recently discussed and compared in detail by Ramanathan \textit{et al.} \cite{Ramanathan2012}. One of the benefits pointed out was the possibility of normalizing a resulting experimental measurement to a nondestructive measurement on the initial sample to overcome the effects of number fluctuations \cite{McKenzie2002}. Based on the result presented in this Brief Report, we argue that it should be possible to employ our scheme to this effect by normalizing a conventional destructive absorption image to that of simple prior dispersive measurement of the peak column density of the sample.

We thank A.~Rakonjac and S.~Hoinka for their contribution to the design and construction of the apparatus. This work was supported by FRST contract NERF-UOOX0703 and the Marsden Fund of New Zealand (Contract No. UOO1121).

\bibliographystyle{apsrev4-1}
%
\end{document}